\documentclass[ letterpaper, amsmath,amssymb, aps,prl,reprint,superscriptaddress, showpacs]{revtex4-1}
\usepackage{graphicx}%
\usepackage{dcolumn}%
\usepackage{bm}%

\newcommand{\ket}[1]{\ensuremath{|{#1}\rangle}}

\begin{document}

\title{Zitterbewegung, Bloch Oscillations and Landau-Zener Tunneling in a Quantum Walk}

\author{Alois Regensburger}

\affiliation{Institute of Optics, Information and Photonics, University Erlangen-Nuremberg, 91058 Erlangen, Germany}
\affiliation{Max Planck Institute for the Science of Light, 91058 Erlangen, Germany}

\author{Christoph Bersch}
\affiliation{Institute of Optics, Information and Photonics, University Erlangen-Nuremberg, 91058 Erlangen, Germany}
\affiliation{Max Planck Institute for the Science of Light, 91058 Erlangen, Germany}

\author{Benjamin Hinrichs}
\affiliation{Institute of Optics, Information and Photonics, University Erlangen-Nuremberg, 91058 Erlangen, Germany}
\affiliation{Max Planck Institute for the Science of Light, 91058 Erlangen, Germany}

\author{Georgy Onishchukov}
\affiliation{Max Planck Institute for the Science of Light, 91058 Erlangen, Germany}

\author{Andreas Schreiber}
\affiliation{Max Planck Institute for the Science of Light, 91058 Erlangen, Germany}

\author{Christine Silberhorn}
\affiliation{Max Planck Institute for the Science of Light, 91058 Erlangen, Germany}
\affiliation{University of Paderborn, Applied Physics, 33098 Paderborn, Germany}

\author{Ulf Peschel}%
\email{ulf.peschel@mpl.mpg.de}%
\affiliation{Institute of Optics, Information and Photonics, University Erlangen-Nuremberg, 91058 Erlangen, Germany}%

\date{\today}

\begin{abstract}
We experimentally investigate a discrete time quantum walk in a system of coupled fiber loops and observe typical phenomena known from the wave propagation in periodic structures as ballistic spreading or an oscillation between two internal quantum states similar to Zitterbewegung (trembling motion). If a position-dependent phase gradient is applied we find localization and Bloch oscillations of the field for moderate as well as Landau-Zener tunneling for strong phase gradients.
\end{abstract}

\maketitle

Compared to the diffusive behavior of a classical random walk, its wave-like analog allows for a ballistic spreading \cite{Aharonov1993a} resulting potentially in an exponential speed-up in quantum algorithms \cite{Childs2009a}. In addition, the so-called quantum walk (QW) seems to be a perfect model system to understand the coherent hopping of excitations between different sites thus explaining basic mechanisms of step-wise energy transfer in e.g. biological systems including photosynthesis \cite{Mohseni2008a,*Plenio2008}. 

QWs have been realized experimentally in several systems, as with trapped atoms \cite{Karski2009a} and ions \cite{Schmitz2009}, photons jumping between different energy levels of a laser cavity \cite{Bouwmeester1999} or propagating in beam splitter arrays \cite{Broome2010} and coupled ring cavities \cite{Schreiber2010}. In all cases the quantum walker travels on a 1D grid and carries an internal two-level quantum state denoted by $\ket{\uparrow}$ and $\ket{\downarrow}$. In each step a unitary operation is applied to this internal state corresponding to tossing a quantum coin. Afterwards the walker either steps to the right or to the left according to its internal state \cite{Kempe2003}. This general scheme can be extended by applying position-dependent unitary operations as phase shifts to the walker's quantum state \cite{Wojcik2004,Schreiber2011}. 

Recently, it was demonstrated that basic features of a QW are also reproduced by photons during their propagation in waveguide arrays \cite{Perets2008} and genuine quantum correlations could be observed in these structures \cite{Peruzzo2010,Bromberg2009}. 

Waveguide arrays deserve attention, in particular because they are well investigated in so-called classical optics, where only the wave-like nature of light is taken into account, but not its genuine quantum character. However, it's still under debate which role genuine quantum effects such as entanglement or multi-particle correlations play in the context of quantum computation algorithms. For some of them as e.g. Grover's search algorithm, the wave interference and coherence properties govern the dynamics \cite{Kwiat2000}. 

Hence, it is worth studying QW from a wave-like perspective thus benefiting from the experience of classical optics. Because the field propagation in waveguide arrays is continuous, the step-like evolution of a discrete quantum walk also adds some new and interesting flavour to further classical considerations. Obviously there are limiting cases where discrete and continuous evolution merge \cite{Strauch2006,*Childs2009b}. However, an investigation has not yet been performed in general. 

The aim of our paper is to deeper study the wave propagation in a discrete QW setup, while particularly focusing on the step-like nature of its evolution and on the influence of a position-dependent phase modulation. For the first time in QW, we experimentally demonstrate the occurence of classical wave-mechanical phenomena as Zitterbewegung (see review article \cite{Zawadzki2011}), Bloch oscillations and Landau-Zener tunneling which are known from solid state physics.

\begin{figure}
\includegraphics{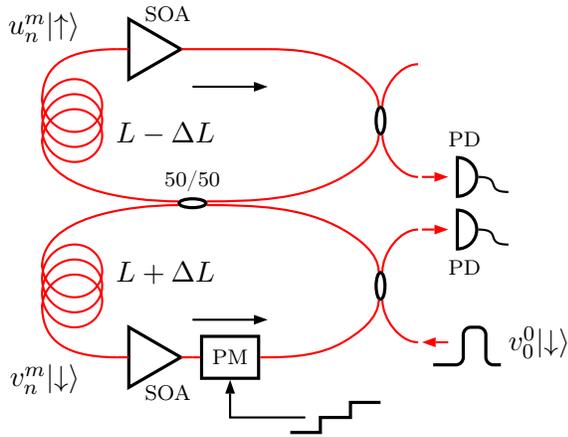}%
\caption{\label{fig:double-loop} Principal scheme of the experimental setup consisting of two fiber loops with length difference $2 \Delta L$ that are connected by a 50/50 coupler; SOA: semiconductor optical amplifiers; PM: phase modulator to introduce position-dependent phase shift $\exp{(i n \alpha)}$; PD: photodiodes.}
\end{figure}

We realize the QW using standard telecommunication equipment in two loops of single-mode fiber with $L=540$~m and a length difference $2 \Delta L=11.4$~m that are coupled by a 50/50 coupler (see Fig. \ref{fig:double-loop}). Coherent light at $1545$~nm is used to simplify the experiments and semiconductor optical amplifiers are inserted to compensate for losses in the setup. As long as the noise generated by the amplifiers remains low, the coherence properties are unaffected and the optical signal will well reproduce the properties of a single quantum particle, but higher order correlations may deviate \cite{Peruzzo2010,Knight2003,*Rohde2011a,*Mayer2010}. A purely passive quantum walk setup could be easily obtained by leaving out the amplifiers and minimizing roundtrip losses.

A rectangular laser pulse with $45$~ns duration is fed into the lower loop simulating a quantum walker starting at position $\ket{0}$. In what follows the two-level internal state of the walker is represented by the distribution of the light field between the two loops corresponding to the states $\ket{\uparrow}$ and $\ket{\downarrow}$. The total state of the system after $m$ steps is represented by 
\begin{equation}
\ket{\Psi(m)} = \sum_n \left( u_n^m \ket{\uparrow} + v_n^m \ket{\downarrow} \right) \otimes \ket{n},
\label{eq:psi_m}
\end{equation}
where $\left|u_n^m \right|^2$ and $\left|v_n^m \right|^2$ are the intensities in the upper and the lower loop for a respective time slot $n$ which are measured with fast photodiodes. During each loop roundtrip the central coupler performs the unbiased coin operation $C= \frac{1}{\sqrt{2}} \left( \begin{smallmatrix} 1 & i \\ i & 1 \end{smallmatrix} \right)$ which acts upon the internal state, where the length difference between the two loops causes a temporal advance or delay thus enabling a step in position space. 

In addition a phase modulator is inserted into the lower loop to further control the evolution. In what follows we restrict to the simplest case---a phase shift which grows linearly with position. Moreover, several additional passive fiber optic components are present in the two loops to control the light's amplitude and polarization and for noise filtering. After each measurement, the residual noise of the optical and electronic amplifiers was recorded in a dark frame and subtracted from the measured signal to improve the signal-to-noise ratio. After averaging over thousands of measurements, the peak intensities of the rectangular pulses were extracted from the data. 

Assuming an exact compensation of losses, the evolution of the amplitudes in the system is well reproduced by the following system of algebraic equations
\begin{equation}
\begin{split}
u_n^{m+1} &= \frac{1}{\sqrt2} \left( u_{n+1}^m + i v_{n+1}^m \right)\\
v_n^{m+1} &= \frac{1}{\sqrt2} \left( v_{n-1}^m + i u_{n-1}^m \right) \exp{ \left( i n \alpha \right)}\,,
\label{eq:rekursion}
\end{split}
\end{equation}
where $\alpha$ is the induced relative phase shift between two positions $n$.
For the notation used above, every second position on the 1D grid is not accessible to the walker. Therefore, the probabilities at these steps are set to 0 in all plots. 

\begin{figure}
\includegraphics[width=\columnwidth]{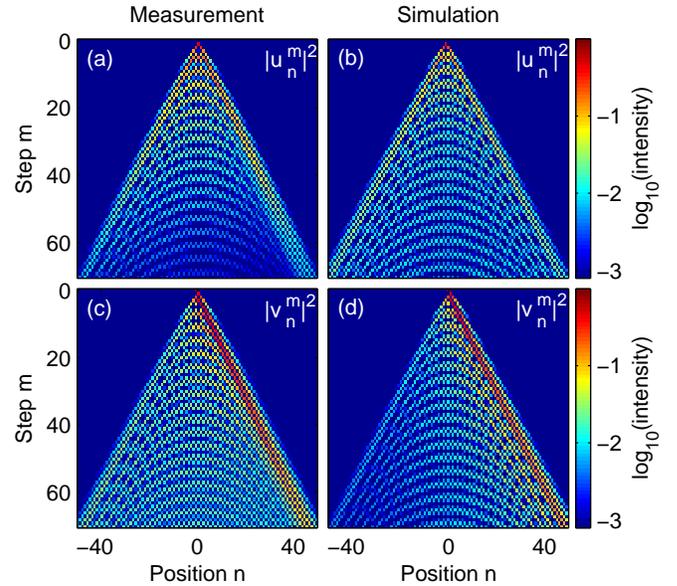}
\caption{\label{fig:qw27_38_expvssim} QW with biased initial state $\ket{\Psi(0)} = \ket{\downarrow} \otimes \ket{0}$ and no phase modulation ($\alpha = 0$); (a),(b): measured and simulated intensities $\left| u_n^m \right|^2$ in the upper loop; (c),(d): measured and simulated intensities $\left| v_n^m \right|^2$ in the lower loop; all in logarithmic scale to enhance visibility.}
\end{figure}

We first analyze the case without phase modulation ($\alpha = 0$). Owing to an exact control of optical gain and polarization, 70 steps of a QW were realized experimentally and a very good agreement with simulations is obtained. In Fig.~\ref{fig:qw27_38_expvssim}, the measured and simulated probability distributions of a quantum walk on the line are displayed. As already mentioned by several authors, the spreading is ballistic and initial asymmetries caused by injecting the initial walker in only one of the loops are conserved. However, we also observe a characteristic pattern which is quite different from conventional diffraction. It shows an oscillation of the energy between the two internal quantum states resulting in a set of nested hyperbolas when displayed on the grid of the QW. 

\begin{figure}
\includegraphics[width=0.7\columnwidth]{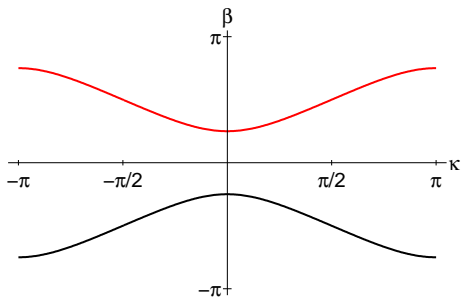}
\caption{\label{fig:bandstrukturqw}Band structure of a QW (equation \ref{eq:dispersionsrelation}); both bands are evenly populated for initial walker state $\ket{\Psi(0)} = \ket{\downarrow} \otimes \ket{0}$; tunneling between bands is possible if force (phase gradient) acts upon the walker.}
\end{figure}

This is the consequence of the two-band nature of the QW, as an analysis of the eigensolutions of Eq. 2 shows. Those are plane waves like $u_n^m=U(\kappa) \exp{\left[ i \beta ( \kappa ) m+i \kappa n \right] }$ and $v_n^m=V(\kappa) \exp{\left[ i \beta ( \kappa ) m+i \kappa n \right] }$ and represent the field distribution in momentum space denoted by $\kappa$ \cite{Pertsch2002}. Both wave vectors $\beta$ and $\kappa$ are $2\pi$ periodic and determine the direction of propagation in the grid of the QW. They are strictly related to each other by \cite{Nayak2000}
\begin{equation}
\beta(\kappa)=\pm \arccos{\left[\frac{1}{\sqrt{2}} \cos(\kappa) \right]}
\label{eq:dispersionsrelation}
\end{equation}
thus forming two mirror-symmetric bands extending in the range $\pi / 4 \leq \left| \beta \right| \leq 3/4 \pi$ (see Fig. \ref{fig:bandstrukturqw}). Hence for each transverse wavenumber $\kappa$ there are two eigenstates with different propagation constants and amplitudes of the internal quantum states $U(\kappa)$ and $V(\kappa)$. This set of eigensolutions is complete and allows to express every initial field distribution. 

For a quantum walker starting at a single point in a fixed internal state ($v_n^0 = \delta_{n0}$ and $u_n^0=0$) all states of the two bands are populated, but run quickly out of phase during evolution. 
For each direction of propagation given by a fixed $\kappa$, this leads to an oscillation of energy between the two internal quantum states denoted by $u$ and $v$. At $\kappa \approx 0$ for waves propagating with low transverse velocity the phase difference between the two bands is  $\Delta \beta = \pi /2$ resulting in an oscillation period of $m=4$, which can easily be recognized in Fig. \ref{fig:qw27_38_expvssim}. Keeping in mind that in the center of the Brillouin zone the double-band structure of the QW is similar to the energy momentum diagram of an electron-positron pair in the Dirac model (see \cite{KURZYNSKI2008,*Strauch2007}) we can interpret these oscillations as an optical manifestation of the Zitterbewegung (trembling motion) \cite{Zawadzki2011} originally proposed by Schr\"odinger when analyzing Dirac's equation.
A quite similar effect was recently observed in a waveguide array with a particularly tailored two element unit cell \cite{Dreisow2010a}. 

In case of the QW, this Zitterbewegung is a generic consequence of the degrees of freedom that originate from the two-level internal state ($\ket{\uparrow}$ and $\ket{\downarrow}$).
For tilted waves the zeros of the Zitterbewegung are shifted because of the curvature of the interacting bands. By approximating the band structure in the center and at the edge of the Brillouin zone by  parabolas [$\beta(\kappa) \approx \pm \left( \frac{\pi}{4} + \frac{\kappa^2}{2} \right)$ and $\beta(\kappa) \approx \mp \left( \frac{3 \pi}{4} + \frac{\left( \kappa - \pi \right)^2}{2} \right)$] and transforming back to position space, one can estimate the field pattern evolving inside the cone of ballistic spreading to be proportional to 
\begin{equation}
\sim \cos{ \left[ \frac{\pi}{4} \left( m - m_0 \right) - \frac{n^2}{2m} \right]}
\label{eq:hyp}
\end{equation}
with $m_0=3$ for $u_n^m$  and $m_0=1$ for $v_n^m$. The resulting set of hyperbolas can be interpreted as an interference pattern of two fields which are subject to diffraction of opposite sign. At $\kappa = \pm \pi/2$ the bandstructure is flat ($\frac{\partial^2}{\partial \kappa^2} \beta(\kappa) = 0$). Respective waves propagate with maximum velocity and do not diffract to first order. This corresponds to ballistic propagation in position space and therefore nicely reflects one of the basic features of a QW. 

As we will see in the following, the particular band structure of QW generates even more surprising effects under the influence of external forces on the system. 

\begin{figure}%
\includegraphics[width=\columnwidth]{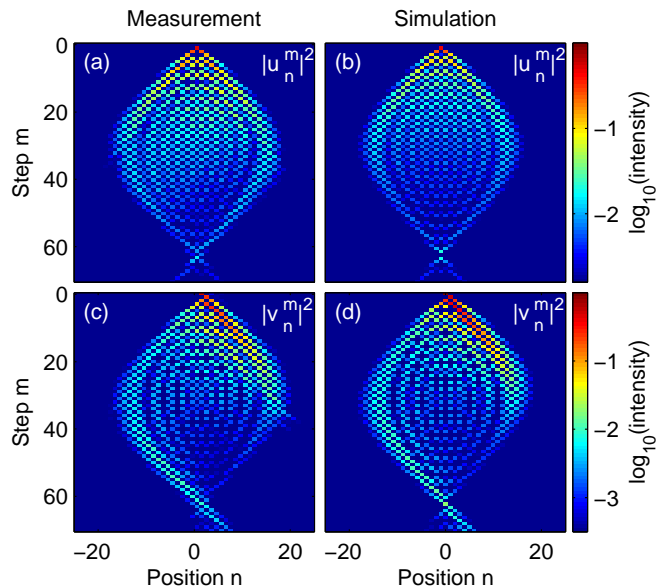}
\caption{\label{fig:bloch20_21_expvssim}QW with biased initial state $\ket{\Psi(0)}= \ket{\downarrow} \otimes \ket{0}$ and moderate position-dependent phase shift ($\alpha = \frac {2\pi}{32}$); fields are displayed as in Fig. \ref{fig:qw27_38_expvssim}.}%
\end{figure}

We focus on the simplest case of a constant force represented by a phase shift ($\alpha \neq 0$) which grows linearly in position $n$ and is applied to the $\ket{\downarrow}$  state at every step $m$. It can be shown that the intensity pattern generated by a quantum walker starting at a single site is only affected by the sum of the phase gradients applied to both internal states. Hence, it does not matter how the phase modulation is distributed between the loops. For a moderate phase gradient $\alpha=2\pi /p$ with $p=32$ we first observe ballistic spreading, which is later stopped and followed by a contraction almost back to the initial state (see Fig. \ref{fig:bloch20_21_expvssim}). The consequence is localization and a periodic recovery \cite{Wojcik2004}. Those so-called Bloch oscillations are known from many areas of physics where wave objects interact with a periodic lattice under the action of a constant force. They were first predicted for electrons in solids subject to an applied electric field and later observed even for photons in biased waveguide arrays \cite{Pertsch1999,Morandotti1999}. 

A simple explanation assumes that the phase gradient induces a constant shift of the field distribution in momentum space by $\delta \kappa = \alpha /2$ in each step. But because Fourier space is periodic in $\kappa$, a recovery must occur after $m=2p$ steps, as it is well reproduced by our experiments.
\begin{figure}%
\includegraphics[width=\columnwidth]{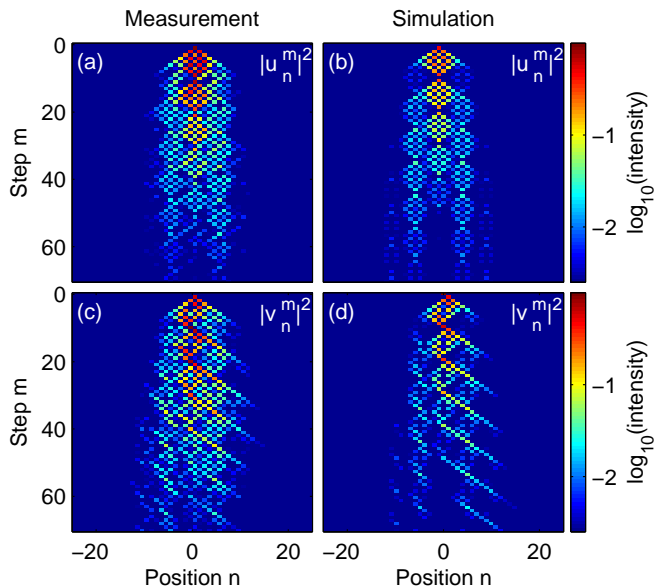}
\caption{\label{fig:bloch34_42_expvssim} QW with biased initial state $\ket{\Psi(0)}= \ket{\downarrow} \otimes \ket{0}$ and strong position-dependent phase shift ($\alpha = \frac {2\pi}{5}$); fields are displayed as in Fig. \ref{fig:qw27_38_expvssim}.}%
\end{figure}%
However, at the outermost line of ballistic spreading, signal power decays like $2^{-m}$, but does not reach zero. Due to the discrete nature of the stepwise evolution, recovery must also be incomplete if $p$ is not an integer or even not rational. However, the resulting deviations are minor. 

The recovery is much more influenced by the magnitude of the phase steps. For higher phase gradient $\alpha$ not only oscillations become faster, but the field starts tunneling to distant lattice sites and recovery is lost (see Fig. \ref{fig:bloch34_42_expvssim}). This phenomenon is again related to the band structure of QW and is known from solid state physics as Landau-Zener tunneling, but has also been observed recently in optical systems \cite{Trompeter2006,Trompeter2006a}. If the phase difference induced between adjacent sites becomes comparable with the spacing between the bands, the field can tunnel. It is important to note that due to the $2\pi$ periodicity of the propagation constant $\beta$ an infinite number of uniformly spaced bands exists and Landau-Zener tunneling can go on forever. In position space tunneling occurs just between lattice sites which acquire the same phase during evolution, i.e. over a distance of roughly $p$ position steps. As a consequence, ballistic spreading is recovered, but at a lower speed.

In summary, we have demonstrated a coherent quantum walk with 70 steps in a system of coupled fiber loops at the telecommunication wavelength. By analyzing the band structure of the system, dynamic features of the QW as e.g. the observed Zitterbewegung can be explained. By implementing a position-dependent coin operator, a discrete analog to Bloch oscillations and Landau-Zener tunneling could be realized. Our experimental system provides a versatile control over the dynamics of photon evolution in a quantum walk and might bring a new perspective to the field of multi-pulsing fiber ring lasers. Moreover, new ideas for quantum algorithms might be derived from the introduced phenomena known from solid state physics. A good scalability to a larger number of steps is expected. 

\begin{acknowledgments}
We acknowledge financial support from DFG Forschergruppe 716, the German-Israeli Foundation and the Cluster of Excellence Engeneering of Advanced Materials (EAM).
\end{acknowledgments}


%

\end{document}